\documentstyle[12pt]{article}

\def\hybrid{\topmargin -20pt    \oddsidemargin 0pt
        \headheight 0pt \headsep 0pt
        \textwidth 6.25in       
        \textheight 9.5in       
        \marginparwidth .875in
        \parskip 5pt plus 1pt   \jot = 1.5ex}
\def\cQ{{\cal Q}}
\def\cG{{\cal G}}
\def\cL{{\cal L}}
\def\cH{{\cal H}}
\def\ket#1{|{#1}\rangle}
\def\noi{\noindent}

\def\baselinestretch{1.2}

\catcode`\@=11

\def\marginnote#1{}
\def\draftlabel#1{{\@bsphack\if@filesw {\let\thepage\relax
   \xdef\@gtempa{\write\@auxout{\string
      \newlabel{#1}{{\@currentlabel}{\thepage}}}}}\@gtempa
   \if@nobreak \ifvmode\nobreak\fi\fi\fi\@esphack}
        \gdef\@eqnlabel{#1}}
\def\@eqnlabel{}
\def\@vacuum{}
\def\draftmarginnote#1{\marginpar{\raggedright\scriptsize\tt#1}}

\def\draft{\oddsidemargin -.2truein
        \def\@oddfoot{\sl preliminary draft \hfil
        \rm\thepage\hfil\sl\today\quad\militarytime}
        \let\@evenfoot\@oddfoot \overfullrule 3pt
        \let\label=\draftlabel
        \let\marginnote=\draftmarginnote
   \def\@eqnnum{(\theequation)\rlap{\kern\marginparsep\tt\@eqnlabel}%
\global\let\@eqnlabel\@vacuum}  }

\def\preprint{\twocolumn\sloppy\flushbottom\parindent 2em
        \leftmargini 2em\leftmarginv .5em\leftmarginvi .5em
        \oddsidemargin -.5in    \evensidemargin -.5in
        \columnsep .4in \footheight 0pt
        \textwidth 10.in        \topmargin  -.4in
        \headheight 12pt \topskip .4in
88      \textheight 6.9in \footskip 0pt
        \def\@oddhead{\thepage\hfil\addtocounter{page}{1}\thepage}
        \let\@evenhead\@oddhead \def\@oddfoot{} \def\@evenfoot{} }

\def\numberbysection{\@addtoreset{equation}{section}
        \def\theequation{\thesection.\arabic{equation}}}

\def\underline#1{\relax\ifmmode\@@underline#1\else
        $\@@underline{\hbox{#1}}$\relax\fi}

\def\titlepage{\@restonecolfalse\if@twocolumn\@restonecoltrue
\onecolumn
     \else \newpage \fi \thispagestyle{empty}\c@page\z@
        \def\thefootnote{\fnsymbol{footnote}} }

\def\endtitlepage{\if@restonecol\twocolumn \else \newpage \fi
        \def\thefootnote{\arabic{footnote}}
        \setcounter{footnote}{0}}  

\catcode`@=12
\relax

\def\figcap{\section*{Figure Captions\markboth
        {FIGURECAPTIONS}{FIGURECAPTIONS}}\list
        {Figure \arabic{enumi}:\hfill}{\settowidth\labelwidth{Figure
999:}
        \leftmargin\labelwidth
        \advance\leftmargin\labelsep\usecounter{enumi}}}
 \relax
\def\tablecap{\section*{Table Captions\markboth
        {TABLECAPTIONS}{TABLECAPTIONS}}\list
        {Table \arabic{enumi}:\hfill}{\settowidth\labelwidth{Table
999:}
        \leftmargin\labelwidth
        \advance\leftmargin\labelsep\usecounter{enumi}}}
 \relax
\def\reflist{\section*{References\markboth
        {REFLIST}{REFLIST}}\list
        {[\arabic{enumi}]\hfill}{\settowidth\labelwidth{[999]}
        \leftmargin\labelwidth
        \advance\leftmargin\labelsep\usecounter{enumi}}}
 \relax
\makeatletter
\newcounter{pubctr}
\def\publist{\@ifnextchar[{\@publist}{\@@publist}}
\def\@publist[#1]{\list
        {[\arabic{pubctr}]\hfill}{\settowidth\labelwidth{[999]}
        \leftmargin\labelwidth
        \advance\leftmargin\labelsep
        \@nmbrlisttrue\def\@listctr{pubctr}
        \setcounter{pubctr}{#1}\addtocounter{pubctr}{-1}}}
\def\@@publist{\list
        {[\arabic{pubctr}]\hfill}{\settowidth\labelwidth{[999]}
        \leftmargin\labelwidth
        \advance\leftmargin\labelsep
        \@nmbrlisttrue\def\@listctr{pubctr}}}
 \relax
\makeatother
\newskip\humongous \humongous=0pt plus 1000pt minus 1000pt

\newif\ifdtup

\font\Scbig=cmss10 scaled\magstep1
\font\Scscr=cmss8 scaled\magstep1
\font\Scscrscr=cmss8
\newfam\Scfam
\textfont\Scfam=\Scbig
\scriptfont\Scfam=\Scscr
\scriptscriptfont\Scfam=\Scscrscr
\def\Sc{\fam\Scfam}

\relax
\hybrid
\def\lvm{\leavevmode\hbox to\parindent{\hfill}}

\def\thefootnote{\fnsymbol{footnote}}
\def\BE{\begin{equation}}
\def\EE{\end{equation}}
\def\BA{\begin{eqnarray}}
\def\EA{\end{eqnarray}}
\def\d{\partial}

\def\th{\theta}

\def\D{\Delta}

\def\tt{\bar\tau}
\def\Gz{\cG_0}
\def\Gn{$\Gz$}
\def\Qz{\cQ_0}
\def\Qn{$\Qz$}

\def\lvm{\leavevmode\hbox to\parindent{\hfill}}
\def\bar{\overline}

\def\L{\left}
\def\R{\right}

\def\BE{\begin{equation}}
\def\EE{\end{equation} \vskip 0.30\baselineskip}
\def\BA{\begin{array}}
\def\EA{\end{array}}

\def\noi{\noindent}

\def\frac#1#2{{\textstyle{{#1}\over{#2}}}}

\def\Kr#1{\delta_{{#1},0}}
\def\ket#1{|{#1}\rangle}

\def\cA{{\cal A}}
\def\cC{{\cal C}}
\def\cG{{\cal G}}
\def\cH{{\cal H}}

\def\cL{{\cal L}}

\def\cQ{{\cal Q}}

\def\cU{{\cal U}}

\def\open#1{\mbox{{\bf{#1}}}}

\def\oZ{{\open Z}}
\def\o1{{\open 1}}

\def\ctop{{\Sc c}}

\def\d{\partial}

\newif\ifold \oldtrue \def\new{\oldfalse}
\let\ssection=\section
\def\section{\setcounter{equation}{0}\ssection}

\begin{document}
\renewcommand{\theequation}{\thesection.\arabic{equation}}
\newcommand{\beq}{\begin{equation}}
\newcommand{\eeq}[1]{\label{#1}\end{equation}}
\newcommand{\ber}{\begin{eqnarray}}
\newcommand{\eer}[1]{\label{#1}\end{eqnarray}}
\begin{titlepage}
\begin{center}

\hfill IMAFF-FM-00/10\\
\hfill NIKHEF-00-004\\
\hfill hep-th/0002081\\
\vskip .5in

{\large \bf Recent Results on N=2 Superconformal Algebras}
\vskip .8in

{\bf Beatriz Gato-Rivera}\\

\vskip .3in

{\em Instituto de Matem\'aticas y F\'\i sica Fundamental, CSIC,\\ Serrano 123,
Madrid 28006, Spain} \footnote{Invited talk presented at the
{\it International Conference on Conformal Field Theory and Integrable
Models}, Landau Institute, Chernogolovka (Russia), June 1999, and at the
{\it 6th International Wigner Symposium}, Bogazici University,
Istambul (Turkey), August 1999.
e-mail address: bgato@imaff.cfmac.csic.es}\\

\vskip .3in

{\em NIKHEF, Kruislaan 409, NL-1098 SJ Amsterdam, The Netherlands}\\

\end{center}

\vskip .6in

\begin{center} {\bf ABSTRACT } \end{center}
\begin{quotation}
In the last six years remarkable developments have taken place concerning
the representation theory of N=2 superconformal algebras. Here we
present the highlights of such developments.
\end{quotation}
\vskip 2cm
February 2000

\end{titlepage}
\vfill
\eject
\def\baselinestretch{1.2}
\baselineskip 16 pt
\section{Introduction to the N=2 Superconformal Algebras}\lvm

\subsection{The N=2 superconformal algebras}
The N=2 superconformal algebras were discovered in the seventies
independently by Ademollo et al. \cite{Ade} and by Kac \cite{Kac}.
The first authors derived the algebras for physical purposes,
in order to define supersymmetric strings, whereas Kac derived
them for mathematical purposes along with his classification of
Lie superalgebras. In modern notation these algebras read

\BE\new\BA{lclclcl}
\L[L_m,L_n\R]&=&(m-n)L_{m+n}+{\ctop\over12}(m^3-m)\Kr{m+n}
\,,&\qquad&[H_m,H_n]&=
&{\ctop\over3}m\Kr{m+n}\,,\\
\L[L_m,G_r^\pm
\R]&=&\L({m\over2}-r\R)G_{m+r}^\pm
\,,&\qquad&[H_m,G_r^\pm]&=&\pm G_{m+r}^\pm\,,
\\
\L[L_m,H_n\R]&=&{}-nH_{m+n}\\
\L\{G_r^-,G_s^+\R\}&=&\multicolumn{5}{l}{2L_{r+s}-(r-s)H_{r+s}+
{\ctop\over3}(r^2-\frac{1}{4})
\Kr{r+s}\,.}\EA\label{N2alg}
\EE

The bosonic generators $L_n$ and $H_n$ correspond to the
stress-energy tensor (Virasoro generators) and to the U(1)
current, respectively. The fermionic generators $G_r^{\pm}$,
with conformal weight 3/2, correspond to the two fermionic
currents. Depending on the modings of the algebra generators
$G_r^{\pm}$ and $H_n$ one distinguishes three N=2 algebras:
the Neveu-Schwarz, the Ramond and the
twisted N=2 algebras. For the Neveu-Schwarz N=2 algebra the
two fermionic generators $G_r^{\pm}$ are half-integer moded
and the U(1) generators $H_n$ are integer moded, i.e.
$r \in {\bf Z} + 1/2, \ \ n \in {\bf Z}$. For the Ramond N=2
algebra $G_r^{\pm}$ and $H_n$ are integer moded, i.e.
$r \in {\bf Z}, \ \ n \in {\bf Z}$ and for the twisted N=2 algebra
$G_r^{+}$ is integer moded whereas $G_s^{-}$ and $H_n$ are
half-integer moded, i.e. $r \in {\bf Z}, \ \ s \in {\bf Z} + 1/2, \ \
n \in {\bf Z} + 1/2 $.

The Neveu-Schwarz, the Ramond and the twisted N=2 superconformal
algebras are not the whole story, however. In 1990 Dijkgraaf,
Verlinde and Verlinde \cite{DVV} presented the Topological
N=2 superconformal algebra, which is the symmetry algebra
of Topological Conformal Field Theory in two dimensions. This
algebra can be obtained from the Neveu-Schwarz N=2 algebra
by `twisting' the stress-energy tensor by addding
the derivative of the U(1) current, procedure known as
`topological twist' \cite{W-top}\cite{EY}. The Topological
N=2 algebra reads

\BE\new\BA{lclclcl}
\L[\cL_m,\cL_n\R]&=&(m-n)\cL_{m+n}\,,&\qquad&[\cH_m,\cH_n]&=
&{\ctop\over3}m\Kr{m+n}\,,\\
\L[\cL_m,\cG_n\R]&=&(m-n)\cG_{m+n}\,,&\qquad&[\cH_m,\cG_n]&=&\cG_{m+n}\,,
\\
\L[\cL_m,\cQ_n\R]&=&-n\cQ_{m+n}\,,&\qquad&[\cH_m,\cQ_n]&=&-\cQ_{m+n}\,,\\
\L[\cL_m,\cH_n\R]&=&\multicolumn{5}{l}{-n\cH_{m+n}+{\ctop\over6}(m^2+m)
\Kr{m+n}\,,}\\
\L\{\cG_m,\cQ_n\R\}&=&\multicolumn{5}{l}{2\cL_{m+n}-2n\cH_{m+n}+
{\ctop\over3}(m^2+m)\Kr{m+n}\,,}\EA\qquad m,~n\in\oZ\,.\label{topalg}
\EE

The fermionic generators $\cG_n$ and $\cQ_n$ , with conformal weight 2
and 1, respectively, and the generators $\cH_n$ are integer moded, i.e.
$n \in {\bf Z}$. The existence of two fermionic zero modes for the Ramond
and for the Topological N=2 algebras allows one to classify the states
in the corresponding Verma modules into two sectors: the $(+)$
and $(-)$ sectors for the Ramond states, $G_0^-$ and $G_0^+$
interpolating between them, and the $\cG$ and $\cQ$ sectors for
the topological states, \Qn\ and \Gn\ interpolating between them.
The two sectors are not the complete description of the states,
however, since there exist indecomposible states in the Verma
modules which do not belong to any of the sectors
\cite{BJI6}\cite{DB1}\cite{DB2}\cite{DB4}. In the case of the Ramond N=2
algebra the indecomposible states are called `no-helicity' states and in
the case of the Topological N=2 algebra they are called `no-label' states.

\subsection{Relations between the N=2 superconformal algebras}
The Neveu-Schwarz and the Ramond N=2 algebras are connected through
the spectral flows. Namely, the even $(\cU_{\theta})$ and the odd
$(\cA_{\theta})$ spectral flows transform the Neveu-Schwarz and the
Ramond N=2 algebras into each other for $\theta={\bf Z}+ 1/2$.
However, they do not map highest weight (h.w.) vectors into h.w.
vectors, but only in particular cases, and they deform
the Verma modules very much, so that they are not isomorphic.

The Neveu-Schwarz and the Topological N=2 algebras are connected
through the topological twists $T_W^{\pm}$. Under these only the
topological h.w. vectors annihilated by \Gn\ are transformed into
h.w. vectors of the Neveu-Schwarz N=2 algebra and the Verma modules
are deformed in a similar way as by the action of the spectral flows for
$\theta= \pm 1/2$. Thus the corresponding Verma modules are not isomorphic.
The topological twists $T_W^{\pm}$ consists of the modification
of the stress-energy tensor by adding the derivative of the U(1)
current: $T(z) \to T(z) \pm 1/2 \ \d H(z)$. As a result the
conformal weights (spins) of the fermionic fields are modified by
$\pm 1/2$, what automatically produces a shift of $\pm 1/2$ in the
modings of these generators: the spin-3/2 fermionic
generators $G_r^+, \ G_r^-$, with $r \in {\bf Z}+ 1/2$, are
transformed into the spin-2 and spin-1 fermionic generators $\cG_n$
and $\cQ_n$, respectively, with $n \in {\bf Z}$. In addition,
$\cQ(z)$ has the properties of a BRST-current, \Qn\ being the BRST
charge. The topological nature of the resulting algebra manifests itself
through the BRST-exactness of the stress-energy tensor:
$\cL_m = 1/2 \{ \cQ_0, \cG_m \}$. When this occurs, the correlators
of the fields of the superconformal field theory do not depend on the
two-dimensional metric, as is well known in the literature (see
for example ref. \cite{Warner}).

Finally, the Ramond and the Topological N=2 algebras are connected
through the composition of the spectral flows with
$\theta={\bf Z}+ 1/2$ and the topological twists. For $\theta=\pm 1/2$
one finds exact isomorphisms between
the states of these algebras level by level \cite{DB4}. The
Verma modules of the Ramond and of the Topological N=2 algebras are
therefore isomorphic.

\section{Representation Theory of the N=2 Superconformal Algebras}

\subsection{Historical overview}
Let us now briefly review the most important developments concerning
the representation theory of the N=2 superconformal algebras. One
can distinguish two periods of remarkable activity. In the first
period, from 1985 until 1988, the determinant formulae for the
Neveu-Schwarz, the Ramond and the twisted N=2 algebras were written
down by various authors, unitarity of the representations was analyzed
and some singular vectors (called simply null vectors) were computed
\cite{BFK}\cite{Nam}\cite{KaMa3}\cite{Yu}\cite{Kir1}\cite{Kir2}\cite{Muss}.
Also some embedding diagrams were presented
\cite{Kir2}\cite{Dobrev}\cite{Matsuo} and the (even) spectral flows
interpolating between the Neveu-Schwarz and the Ramond N=2 algebras were
written down \cite{SS}.

During the period from 1989 until 1993 there was not much activity in
the representation theory of N=2 algebras. However, two
developments took place which were of crucial importance. On one side
the chiral rings were discovered \cite{LVW} and with them one realized
the necessity to analyze the chiral representations of the N=2 algebras.
On the other side the Topological N=2 algebra was written down
\cite{DVV} and its importance for string theory was also realized
\cite{GRS}\cite{BLNW}.

In the second period of remarkable activity, from 1994 until nowadays,
there have been several important findings regarding the states in the
Verma modules of the N=2 algebras: the discovery of two-dimensional
singular spaces \cite{Doerr2}\cite{DB3}, the discovery of subsingular
vectors \cite{BJI5}\cite{BJI6}\cite{BJI7}\cite{DB3} and the discovery
of indecomposible states \cite{BJI6}\cite{DB4}. In addition an
(almost) complete
classification of embedding diagrams was presented for the
Neveu-Schwarz N=2 algebra \cite{Doerrthesis}\cite{Doerr3} and
the determinant formulae for the Topological N=2 algebra were computed
\cite{DB4} as well as the determinant formulae for the chiral
representations of the Topological, the Neveu-Schwarz and the Ramond
N=2 algebras \cite{BJI7}. Moreover, the odd spectral flows were written
down \cite{BJI4}\cite{Be1}, which are believed to provide the
complete set of automorphisms for the N=2 algebras. Furthermore, recently
a powerful tool has been developed for the study of the representations
of any Lie algebra or superalgebra, the so-called `adapted ordering method'.
This method has been applied so far sucessfully to the N=2 algebras and
to the Ramond N=1 algebra. In what follows we will say a few words about
all these new results and developments.

\subsection{The odd spectral flow}
The odd spectral flow $\cA_{\theta}$, when acting on the states and
generators of the Neveu-Schwarz and the Ramond N=2 algebras read
\cite{BJI4}\cite{Be1}

\BE\new\BA{rclcrcl}
\cA_\th \, L_m \, \cA_\th^{-1}&=& L_m
 +\th H_m + {\ctop\over 6} \th^2 \delta_{m,0}\,,\\
\cA_\th H_m \, \cA_\th^{-1}&=&- H_m - {\ctop\over3} \th \delta_{m,0}\,,\\
\cA_\th \, G^+_r \, \cA_\th^{-1}&=&G_{r-\th}^-\,,\\
\cA_\th \, G^-_r \, \cA_\th^{-1}&=&G_{r+\th}^+\,,\
\label{ospfl} \EA\EE

\noi
with $\cA_{\theta}^{-1}=\cA_{\theta}$ (it is therefore an involution).
$\cA_{\theta}$ and the even (usual) spectral flow $\cU_{\theta}$ \cite{SS}
are quasi-mirror symmetric under $H_m \leftrightarrow -H_m, \
G_r^+ \leftrightarrow G_r^-, \ \theta \leftrightarrow -\theta $.
However, $\cA_{\theta}$ generates $\cU_{\theta}$ and consequently it
is the only fundamental spectral flow, as one can see in the composition
rules

\BE  \cU_{\th_2} \ \cU_{\th_1} = \cU_{(\th_2 + \th_1)} , \qquad
\cA_{\th_2} \ \cA_{\th_1} = \cU_{(\th_2 - \th_1)} ,  \EE

\BE  \cA_{\th_2} \ \cU_{\th_1} = \cA_{(\th_2 - \th_1)} , \qquad
\cU_{\th_2} \ \cA_{\th_1} = \cA_{(\th_2 + \th_1)} .  \EE

\noi
$\cA_{\theta}$ is believed to provide the complete set of automorphisms
of the N=2 algebras (for $\theta \in {\bf Z}$), whereas $\cU_{\theta}$
provide only `half' of them.

\subsection{Indecomposible singular vectors}
The indecomposible singular vectors of the Ramond N=2 algebra
were overlooked until very recently (they do not exist for
the Neveu-Schwarz N=2 algebra). In fact, they were discovered
first for the Topological N=2 algebra in ref. \cite{BJI6}, where they
were called `no-label' singular vectors. Shortly afterwards some
examples were presented for the Ramond N=2 algebra \cite{DB1} under the
name `no-helicity' singular vectors.

At first sight, just by inspecting the anticommutator of the fermionic
zero modes, one realizes that indecomposible singular vectors are
allowed to exist by the Topological (and Ramond) N=2 algebra. Namely,
from $\{ \cG_0, \cQ_0 \} = 2 \cL_0$ one deduces that for non-zero
conformal weight $\Delta \neq 0$ all the states can be decomposed into
linear combinations of \Gn-closed states and \Qn-closed states
(i.e. states annihilated by \Gn\ and states annihilated by \Qn ):

\BE \ket{\chi}={1\over 2 \D}(\Gz \Qz \ket{\chi}+\Qz \Gz \ket{\chi})\,.\EE

For zero conformal weight $\Delta = 0$, however, there is not such a
decomposition and the states can be annihilated either by one of the
fermionic zero modes or by both or by none of them. The latter are the
indecomposible `no-label' states. In the case of the Ramond N=2 algebra,
from the anticommutator $\{ G_0^+, G_0^- \} = 2 L_0 - {\ctop \over 12}$
one deduces that the `no-helicity' indecomposible states are allowed
for $\Delta = {\ctop \over 24}$.

Indecomposible singular vectors are also allowed by the analysis of
maximal dimensions \cite{DB2} (see later). They actually exist
already at levels 1 and 2 \cite{BJI6}\cite{DB1} and recently it has
been proved that they {\it must} necessarily exist \cite{DB4}. The
argument goes as follows. One analyzes two curves in the parameter
space of singular vectors. Each curve corresponds to two different families
of singular vectors. In some discrete intersection points, however,
the four singular vectors reduce to three (two of them coincide).
But the rank of the inner product matrix is upper semi-continuous
and therefore another singular vector must exist. By analyzing the
dimensions of the singular vectors involved one deduces finally that
the new singular vector must be indecomposible.

\subsection{New determinant formulae}
The determinant formulae for the Topological N=2 algebra have been
computed for generic (standard) Verma modules, for chiral Verma
modules and for no-label Verma modules \cite{BJI7}\cite{DB4}.
In addition, determinant formulae have been computed for the chiral Verma
modules of the Neveu-Schwarz and of the Ramond N=2 algebras \cite{BJI7},
as well as for the no-helicity Verma modules of the Ramond N=2 algebra
(as a straightforward derivation of the results for the no-label Verma
modules of the Topological N=2 algebra) \cite{DB4}.

The generic Verma modules of the Topological N=2 algebra are built
on \Gn-closed and/or \Qn-closed h.w. vectors (annihilated either
by $\cG_0$ or by $\cQ_0$). For conformal weight $\Delta \neq 0$ there
are two h.w. vectors at the bottom of the Verma modules (one is
\Gn-closed and the other \Qn-closed), giving rise to the two sectors:
the $\cG$-sector and the $\cQ$-sector. For $\Delta = 0$, however,
there is only one h.w. vector (plus one singular vector at level zero)
at the bottom of the Verma modules.

The chiral Verma modules of the Topological N=2 algebra are built on
chiral h.w. vectors annihilated by both \Gn\ and \Qn . Therefore they
have only one h.w. vector at the bottom, which has zero conformal weight
$\Delta = 0$. They are incomplete Verma modues that can be realized as
quotient modules of a generic Verma module with $\Delta = 0$ divided
by the submodule generated by its level zero singular vector. Similarly,
the chiral Verma modules of the Ramond N=2 algebra are built on h.w.
vectors annihilated by both $G_0^+$ and $G_0^-$, with $\D={\ctop\over24}$.
The chiral (anti-chiral) Verma modules of the Neveu-Schwarz N=2 algebra,
in turn, are built on chiral (anti-chiral) h.w. vectors annihilated by
$G^+_{-1/2}$ ($G^-_{-1/2}$), satisfying $\D = {h \over2}$
($\D={-h\over2}$), where $h$ is the U(1) charge.

Finally, no-label (no-helicity) Verma modules are built on no-label
(no-helicity) h.w. vectors. At the bottom they consist of one h.w.
vector plus three singular vectors at level zero obtained by the action
of \Gn\ and \Qn\ ($G_0^+$ and $G_0^-$) on the no-label (no-helicity) h.w.
vector. No-label (no-helicity) Verma modules {\it appear as submodules}
inside generic Verma modules, the bottom of these submodules consisting
of four singular vectors consequently.

Apart from no-label (no-helicity) submodules, in generic Verma modules
of the Topological (Ramond) N=2 algebra one can find
another three types of submodules, taking into account the size and
the shape at the bottom of the submodule \cite{DB4}. In chiral Verma
modules, however, one finds only one kind of submodule
\cite{BJI7}\cite{DB4}.

\subsection{Discovery of subsingular vectors}
Subsingular vectors are null, but not h.w. vectors, that become singular
(i.e. h.w. vectors) after the quotient of the Verma module by a
submodule. That is, they become singular by setting one singular
vector to zero. This implies that they are located {\it outside} that
submodule since otherwise they would go away after the quotient.

Subsingular vectors for the Neveu-Schwarz, the Ramond and the Topological
N=2 algebras were reported for the first time in 1996 - 1997 in refs.
\cite{BJI5}\cite{BJI6}\cite{BJI7}. For the twisted N=2 algebra subsingular
vectors were presented
only recently in ref. \cite{DB3}. The discovery of subsingular
vectors for the N=2 algebras was as follows. From the chiral determinant
formulae for the Neveu-Schwarz, the Ramond and the Topological N=2 algebras
\cite{BJI7} one deduces the existence of singular vectors in the
chiral Verma modules that are not singular in the complete (generic)
Verma modules before the quotient that gives rise to the chiral Verma
modules. These are therefore subsingular vectors in the generic Verma
modules. Many explicit examples of subsingular vectors of this type
(i.e. becoming singular in the chiral Verma modules) were presented
for the Neveu-Schwarz, the Ramond and the Topological N=2 algebras
\cite{BJI6}\cite{BJI7}.
For the twisted N=2 algebra subsingular vectors were found by analyzing the
Verma modules using the `adapted ordering method' \cite{DB3} (see later).

\subsection{Discovery of two-dimensional singular vector spaces}
The two-dimensional singular vector spaces were discovered first for the
Neveu-Schwarz N=2 algebra by D\"orrzapf \cite{Doerr2}\cite{Doerrthesis}
in 1994.
In particular, some conditions were found to guarantee the existence
of two-dimensional spaces of {\it uncharged} singular vectors. These
conditions can be written as the simultaneous vanishing of two functions:
$\epsilon_r^+(h,\ctop) = \epsilon_r^-(h,\ctop) = 0$,
where $h$ is the U(1) charge
of the h.w. vector of the Verma module. The corresponding uncharged
singular vectors are located in Verma modules where two {\it charged}
and one {\it uncharged} singular vectors intersect (at different levels)
in such a way that the charged singular vectors are the primitive ones
and the two-dimensional uncharged singular space is
secondary of the two charged
singular vectors. These two-dimensional spaces are spanned by a tangent
space of vanishing surfaces corresponding to singular vectors. Once the
`general formula' for singular vectors vanish one gets the
two-dimensional singular space in the tangent space.

An straightforward extension of these results to the Topological N=2
algebra was presented in ref. \cite{BJI6}. As a result four types of
topological singular vectors were found for which two-dimensional
spaces may exist. The extension of these results to the Ramond N=2
algebra is also straightforward since the corresponding Verma modules
are isomorphic to the Verma modules of the Topological N=2 algebra.
For the twisted N=2 algebra two-dimensional singular spaces were found
using the adapted ordering method \cite{DB3}. In this case, however,
the two-dimensional singular spaces are of different nature than the
ones corresponding to the other N=2 algebras: they are spanned by two
{\it primitive} singular vectors instead of two secondary singular vectors.

\subsection{N=2 Embedding diagrams}
In 1995 D\"orrzapf presented a complete classification of embedding
diagrams for the Neveu-Schwarz N=2 algebra \cite{Doerrthesis}
(see also ref. \cite{Doerr3}). He
proved that the relative charge $q$ of all singular vectors (not
only the primitive ones) satisfy $|q| \leq 1$, correcting many earlier
diagrams in the literature \cite{Kir2}, and he presented many more
diagrams than previously known \cite{Dobrev}\cite{Matsuo}. These
results have not been improved so far although we know that they must
be improved because subsingular vectors were assumed not to exist,
being discovered one year afterwards.

The embedding diagrams of the Neveu-Schwarz N=2 algebra can be carefully
adapted to provide embedding diagrams for the Topological and for the
Ramond N=2 algebras. One has to take into account, however, that many
singular vectors of these algebras do not correspond to singular vectors
of the Neveu-Schwarz N=2 algebra but to null descendants of singular
vectors or even to subsingular vectors (for example, indecomposable
no-label and no-helicity singular vectors of the Topological and of the
Ramond N=2 algebras always correspond to subsingular vectors of the
Neveu-Schwarz N=2 algebra \cite{DB1}).

\section{The Adapted Ordering Method}
The `adapted ordering method', developed in ref. \cite{DB2}, can
be applied to most Lie algebras and superalgebras. It allows:

i) to determine {\it maximal dimensions} for a given type of singular
   vector space,

ii) to rule out the existence of certain types of singular vectors
(with dimension zero),

iii) to identify all singular vectors by only a few coefficients,

iv) to spot subsingular vectors,

v) to obtain easily product expressions of singular vector operators
in order to compute secondary singular vectors (or decide whether they
vanish),

vi) to set the basis for constructing embedding diagrams.

The method originates (in rudimentary form) from a procedure developed
by Kent for the analytically continued Virasoro algebra \cite{Kent}.
The analytical continuation is not necessary,
however, for the adapted ordering method. The key idea of this method
is to find a suitable ordering for the terms of the singular vectors, i.e.
a criterion to decide which of two terms is the bigger one, for example
between the terms at level 4 and charge 1:  $\ G_{-2}^+ L_{-2}\ $ and
$\ G_{-1}^+ H_{-2}L_{-1}$. The ordering must be {\it adapted} to a
subset of terms $\cC^A_{l,q} \in \cC_{l,q}$, where $\cC_{l,q}$ is the
set of all possible terms at level $l$ with charge $q$ (for the details
see ref. \cite{DB2}).

The crucial point of this method is the following. The complement of
$\cC^A_{l,q}$ is the ordering kernel
$\cC^K_{l,q} = \cC_{l,q} / \cC^A_{l,q}$ and its size puts a limit on the
dimension of the corresponding singular vector space. Namely, if the
ordering kernel $\cC^K_{l,q}$ has $n$ elements then there are at
most $n$ linearly independent singular vectors $\Psi_{l,q}$, at
level $l$ with charge $q$, in a given Verma module. In other words,
the singular vectors $\Psi_{l,q}$ span a singular space that is at
most $n$-dimensional. As a result, if $\cC^K_{l,q} = \not0$ then
there are no singular vectors of type $\Psi_{l,q}$ (the singular space
has dimension zero). Therefore we need to find a suitable, clever
ordering in order to obtain the smallest possible kernel. Furthermore,
the coefficients with respect to the terms of the ordering kernel
uniquely identify a singular vector. This implies that just a few
(one, two,...) coefficients completely determine a singular vector
no matter its size. As a consequence one can find easily product
expressions for descendant singular vectors and set the basis to
construct embedding diagrams.

The adapted ordering method has been applied to the Topological, to the
Neveu-Schwarz and to the Ramond N=2 algebras in ref. \cite{DB2} and to
the twisted N=2 algebra in ref. \cite{DB3}. The maximal dimensions
of the existing types of singular vectors have been found to be one
or two, with the exception of some types of singular vectors in
`no-label' and `no-helicity' Verma modules for which the maximal
dimension has been found to be three. For the Topological and the Ramond
N=2 algebras the only existing types of singular vectors (primitive as
well as secondary), distinguished by the relative charge $q$ and the
annihilation properties under the fermionic zero modes, resulted in:
twenty types in generic Verma modules, with $|q| \leq 2$, nine types
in no-label and no-helicity Verma modules, with $|q| \leq 2$, and
four types in chiral Verma modules, with $|q| \leq 1$. These results
had been conjectured previously in ref. \cite{BJI6}. For the
Neveu-Schwarz N=2 algebra one obtained $|q| \leq 1$, in agreement with
the results presented in ref. \cite{Doerr2}. As
we pointed out before, in the case of the twisted N=2 algebra the
application of the adapted ordering method has lead to the discovery
of subsingular vectors and two-dimensional singular spaces for this
algebra.

\section{Final Remarks}
In spite of the progress made in the last six years, the representation
theory of the N=2 superconformal algebras is not finished yet. It remains
to classify the subsingular vectors and to complete the classification
of embedding diagrams. Several of the techniques that have been used for
the analysis of these algebras can be easily transferred to the analysis
of other Lie algebras and superalgebras. This holds especially for
the adapted ordering method, that has already
been applied sucessfully to the Ramond N=1 superconformal algebra
\cite{Doerr4}, leading as a result to the discovery of two-dimensional
singular spaces and subsingular vectors, which do not exist for the
Neveu-Schwarz N=1 superconformal algebra.

\vskip 1.5cm

\centerline{\bf Acknowledgements}

I am very grateful to Prof. Alexander Belavin for the invitation to
participate in the {\it Conference on Conformal Field Theory and
Integrable Models}, at Landau Institute, and for his hospitality. Also
I would like to thank the organizers of the {\it 6th International
Wigner Symposium}, at Bogazici University, for the invitation to
participate.


\begin{thebibliography}{9}
\def\NPB{Nucl. Phys. B}
\def\PLB{Phys. Lett. B}
\def\MPLA{Mod. Phys. Lett. A}

\bibitem{Ade} M. Ademollo et al., \PLB62 (1976) 105;
 \NPB111 (1976) 77; \NPB114 (1976) 297;

\bibitem{Kac} V.G. Kac, {\it Lie superalgebras}, Adv. Math. 26:8, (1977)


\bibitem{DVV} R. Dijkgraaf, E. Verlinde and H. Verlinde, \NPB352
(1991) 59

\bibitem{W-top} E.~Witten, Commun. Math. Phys. 118 (1988) 411;
 \NPB340 (1990) 281

\bibitem{EY} T.~Eguchi and S.~K.~Yang, \MPLA5 (1990) 1653

\bibitem{Warner} N.P. Warner, {\it N=2 Supersymmetry, Integrable Models and
Topological Field Theories}, Trieste lectures, hep-th/9301088 (1993)

\bibitem{BFK} W. Boucher, D. Friedan and A. Kent, {\it Determinant Formulae
and Unitarity for the N=2 Superconformal Algebras in two Dimensions or
Exact Results on String Compactification},
Phys. Lett. B172 (1986) 316

\bibitem{Nam} S. Nam, \PLB172 (1986) 323

\bibitem{KaMa3} M. Kato and S. Matsuda, {\it Null Field Construction and
Kac Formulae of N=2 Superconformal Algebras in two Dimensions},
\PLB184 (1987) 184

\bibitem{Yu} P. Di Vecchia, J.L. Petersen and M. Yu, \PLB172 (1986) 211; \\
P. Di Vecchia, J.L. Petersen, M. Yu and H.B. Zheng \PLB174 (1986) 280; \\
M. Yu and H.B. Zheng, \NPB288 (1987) 275

\bibitem{Kir1} E.B. Kiritsis, {\it Structure of N=2 Superconformally
Invariant Unitary Minimal Theories: Operator Algebra and Correlation
Functions}, Phys. Rev. D36 (1987) 3048

\bibitem{Kir2} E.B. Kiritsis, {\it Character Formulae and the Structure
of the Representations of the N=1, N=2 Superconformal Algebras},
Int. J. Mod. Phys A3 (1988) 1871

\bibitem{Muss} G. Mussardo, G. Sotkov and M. Stanishkov, {\it N=2
Superconformal Minimal Models}, Int. J. Mod. Phys. A4 (1989) 1135

\bibitem{Dobrev} V.K. Dobrev, {\it Characters of the Unitarizable Highest
Weight Modules over the N=2 Superconformal Algebras},
\PLB186 (1987) 43

\bibitem{Matsuo} Y. Matsuo, {\it Character Formula of the $c<1$ Unitary
Representation of the N=2 Superconformal Algebra},
Prog. Theor. Phys. 77 (1987) 793

\bibitem{SS} A. Schwimmer and N. Seiberg, {\it Comments on the N = 2, 3, 4
Superconformal Algebras in Two Dimensions}, \PLB184 (1987) 191

\bibitem{LVW} W.~Lerche, C.~Vafa and N.~P.~Warner,
 \NPB324 (1989) 427

\bibitem{GRS} B.~Gato-Rivera and A.~M.~Semikhatov, \PLB293 (1992) 72,
 Theor. Mat. Fiz. 95 (1993) 239, Theor. Math. Phys. 95 (1993) 536;
 \NPB408 (1993) 133

\bibitem{BLNW} M. Bershadsky, W. Lerche, D. Nemeschansky and
N.P. Warner, \NPB401 (1993) 304

\bibitem{BJI4} B. Gato-Rivera and J.I. Rosado, {\it The Other Spectral Flow},
\MPLA11 (1996) 423, hep-th/9511208

\bibitem{Be1} B. Gato-Rivera, {\it The Even and the Odd Spectral Flows on
the N=2 Superconformal Algebras},  Nucl. Phys. B512 (1998) 431,
hep-th/9707211

\bibitem{BJI5} B. Gato-Rivera and J.I. Rosado, {\it Interpretation of the
Determinant Formulae for the Chiral Representations of the N=2
Superconformal Algebra}, IMAFF-96/38, NIKHEF-96-007, hep-th/9602166 (1996)

\bibitem{BJI6} B. Gato-Rivera and J.I. Rosado, {\it Families of Singular
and Subsingular Vectors of the Topological N=2 Superconformal
Algebra}, Nucl. Phys. B514 [PM] (1998) 477, hep-th/9701041

\bibitem{BJI7} B. Gato-Rivera and J.I. Rosado, {\it Chiral Determinant
Formulae and Subsingular Vectors for the N=2 Superconformal Algebras},
 \NPB503 (1997) 447, hep-th/9706041

\bibitem{Doerr2} M.~D\"orrzapf, {\it Analytical Expressions for Singular
Vectors of the N=2 Superconformal Algebra},
Commun. Math. Phys. 180 (1996) 195

\bibitem{Doerrthesis} M.~D\"orrzapf, {\it Superconformal Field Theories
and their Representations}, Ph. D. thesis, University of Cambridge,
September 1995,\\
http://www.damtp.cam.ac.uk/user/md131/research/thesis.html

\bibitem{Doerr3} M.~D\"orrzapf, {\it The embedding structure of unitary
N = 2 minimal models}, \NPB529 (1998) 639, hep-th/9712165.

\bibitem{Doerr4} M.~D\"orrzapf, {\it Highest weight representations
of the N = 1 Ramond algebra}, DAMTP-99-28 (1999), hep-th/9905150.


\bibitem{DB1} M. D\"orrzapf and B. Gato-Rivera, {\it Transmutations between
Singular and Subsingular Vectors of the N=2 Superconformal Algebras},
\NPB557 [PM] (1999) 517, hep-th/9712085

\bibitem{DB2} M. D\"orrzapf and B. Gato-Rivera, {\it Singular Dimensions of
the N=2 Superconformal Algebras.I}, Commun. Math. Phys. 206, 493 (1999),
hep-th/9807234

\bibitem{DB3} M. D\"orrzapf and B. Gato-Rivera, {\it Singular Dimensions of
the N=2 Superconformal Algebras.II. The Twisted N=2 Algebra}, DAMTP 99-19,
IMAFF-FM-99/08, hep-th/9902044 (1999)

\bibitem{DB4} M. D\"orrzapf and B. Gato-Rivera, {\it Determinant Formulae
for the Topological N=2 Superconformal Algebra},
\NPB558 [PM], 503 (1999), hep-th/9905063

\bibitem{Kent} A. Kent, {\it  Singular Vectors of the Virasoro Algebra},
\PLB273 (1991) 56

\end{thebibliography}
\end{document}